\documentclass[aps,pra,showpacs,twocolumn,amsmath,amssymb,superscriptaddress]{revtex4}

\usepackage[english]{babel}
\usepackage{latexsym}
\usepackage{graphics}
\usepackage{epsfig}
\usepackage{color}
\usepackage{url}
\usepackage[normalem]{ulem}

\def\be{\begin{equation}}
\def\ee{\end{equation}}
\def\bea{\begin{eqnarray}}
\def\eea{\end{eqnarray}}
\def\bi{\begin{itemize}}
\def\ei{\end{itemize}}
\def\bin{\begin{enumerate}}
\def\ein{\end{enumerate}}

\newcommand{\vect}[1]{\mathbf{#1}}

\usepackage{graphicx}
\begin{document}
\title{Condensate Phase Microscopy}


\author{Arkadiusz Kosior}
\affiliation{
Instytut Fizyki imienia Mariana Smoluchowskiego,
Uniwersytet Jagiello\'nski, ulica Reymonta 4, PL-30-059 Krak\'ow, Poland}

\author{Krzysztof Sacha}
\affiliation{
Instytut Fizyki imienia Mariana Smoluchowskiego,
Uniwersytet Jagiello\'nski, ulica Reymonta 4, PL-30-059 Krak\'ow, Poland}
\affiliation{
Mark Kac Complex Systems Research Center, 
Uniwersytet Jagiello\'nski, ulica Reymonta 4, PL-30-059 Krak\'ow, Poland}

\date{\today}

\begin{abstract}
We show that the phase of a Bose-Einstein condensate wave-function of ultra-cold atoms in an optical lattice potential in two-dimensions can be detected. The time-of-flight images, obtained in a free expansion of initially trapped atoms, are related to the initial distribution of atomic momenta but the information on the phase is lost. However, the initial atomic cloud is bounded and this information, in addition to the time-of-flight images, is sufficient in order to employ the phase retrieval algorithms. We analyze the phase retrieval methods for model wave-functions  in a case of a Bose-Einstein condensate in a triangular optical lattice in the presence of artificial gauge fields.
\end{abstract}


\maketitle

The development of experimental techniques in ultra-cold atomic gases is tremendous \cite{exptech}. Both trapping potentials for atoms and mutual atom interactions can be controlled and engineered. The artificial magnetic fields or even the non-abelian gauge potentials that are experienced by the neutral particles can be created in a laboratory \cite{gauge}. Similar rapid progress is observed in detection methods. Nowadays, distributions of atoms can be measured even with a single atom resolution \cite{greiner08,greiner10,bloch10}.  

A typical Bose-Einstein condensate (BEC) of ultra-cold atoms can be described by a real condensate wave-function (order parameter) and the appearance of a non-uniform phase of the condensate reflects thermal fluctuations only \cite{lewenstein01,arlt03}. However, in the presence of artificial gauge potentials or in multi-orbital superfluid phase of ultra-cold atoms in optical lattices, the order parameter may become intrinsically complex  \cite{struck2011,hemmerich11,sengstock12,spielman12,olschlager2013}. In these cases, extracting information on the phase of the condensate would be invaluable. In particular, one would be able to reconstruct Berry's phases related to synthetic abelian gauge fields.  For example, the fully frustrated triangular lattice problem reveals a doubly degenerate ground state which corresponds to two distinct configurations  on the $\pi-$flux lattice \cite{struck2011,struck2013}. In such a system there is a theoretical possibility of spatial domains formation where different 
parts of the 
system are in two different groundstate 
configurations. Information on the condensate phase would allow experimentalists to reveal the domains (for experiments with ferromagnetic domain formation see \cite{parker13}).

Atomic density measurement after time-of-flight (TOF) is a standard detection technique in ultra-cold atoms expierments \cite{exptech}. For sufficiently long TOF the density images reflect the initial distributions of atomic momenta. The distributions alone are not sufficient to invert the Fourier transform and obtain the condensate wave-function in the configuration space as the phase information is lost. However, we will show that the phase of the condensate wave-function can be reconstructed if a system is initially bounded, which is always the case for trapped atoms. 
In crystallography, electron microscopy and astronomical imaging, computationally retrieving the phase of diffraction patterns is remarkably successful \cite{Fienup1987,miaosayre2000,marchesini2007}. The examples range from the biological cells imaging to the evaluation of the aberrations in the Hubble space telescope \cite{marchesini2007}. On the other hand, in the present letter it is not an external wave that diffracts on a measured object and is subsequently detected, but the matter wave itself is the {\it object} to be reconstructed.

In the following we concentrate on atomic gases prepared in an optical lattice potential \cite{exptech}. After a sudden turn off of the potential and a period of nearly free expansion, the atomic density replicates the initial momentum distribution of the system. In the far field regime, i.e. for a very long time-of-flight $t_{TOF}$, the atomic density detected by a CCD camera reads
\be
I(\vect{r}) \propto |\tilde\psi_0(\vect{k})|^2\propto \left|\sum_i e^{i \vect{k}\cdot\vect{r}_i}\psi(\vect{r}_i)\right|^2, \quad \vect{k}=\frac{m \vect{r}}{\hbar \, t_{TOF}},
\label{intensity}
\ee
where $\psi(\vect{r}_i)$ is the initial condensate wave-function, $\vect{r}_i$'s denote positions of the lattice sites and $m$ is the atomic mass. Multidimensional system of equations (\ref{intensity}), quadratic in $\psi(\vect{r}_i)$, is difficult to solve. Phase retrieval algorithms try to find the condensate wave-function iteratively by imposing some extra conditions such as a finite support $S$ of $\psi(\vect{r}_i)$ \cite{Fienup1987,miaosayre2000,marchesini2007}. The momentum distribution, i.e. the squared modulus of the Fourier transform of the object $\psi(\vect{r}_i)$, is known. Consequently, the sought wave-function belongs to a set of all objects with the same modulus of its Fourier transform $|\tilde{\psi}_0(\vect{k})|$. It also belongs to a set of objects with the same support $S$. Phase retrieval algorithms seek for the intersection of these two sets, where the solution is located. In a basic version of the algorithm, starting with a randomly chosen $\psi^{(0)}(\vect r)$ that is localized on 
$S$, the following procedure is applied: (i) the Fourier transform is used resulting in $|\tilde\psi^{(0)}(\vect k)|e^{i\tilde\varphi^{(0)}(\vect k)}$, (ii) $|\tilde\psi^{(0)}(\vect k)|$ is substituted with $|\tilde\psi_0(\vect{k})|$, Eq.~(\ref{intensity}), and the inverse Fourier transform is applied, (iii) a new $\psi^{(1)}(\vect r)$ is obtained by setting zeros in the result of the inverse transform for $\vect r \notin S$ and the entire procedure is repeated until the wave-function converges. If additional information, e.g. the modulus $|\psi(\vect{r})|$, is available, it can be used in the phase retrieval algorithms, see \cite{supplement} for all technical details.

\begin{figure}[ht]
\begin{center}
\resizebox{0.5\columnwidth}{!}{\includegraphics{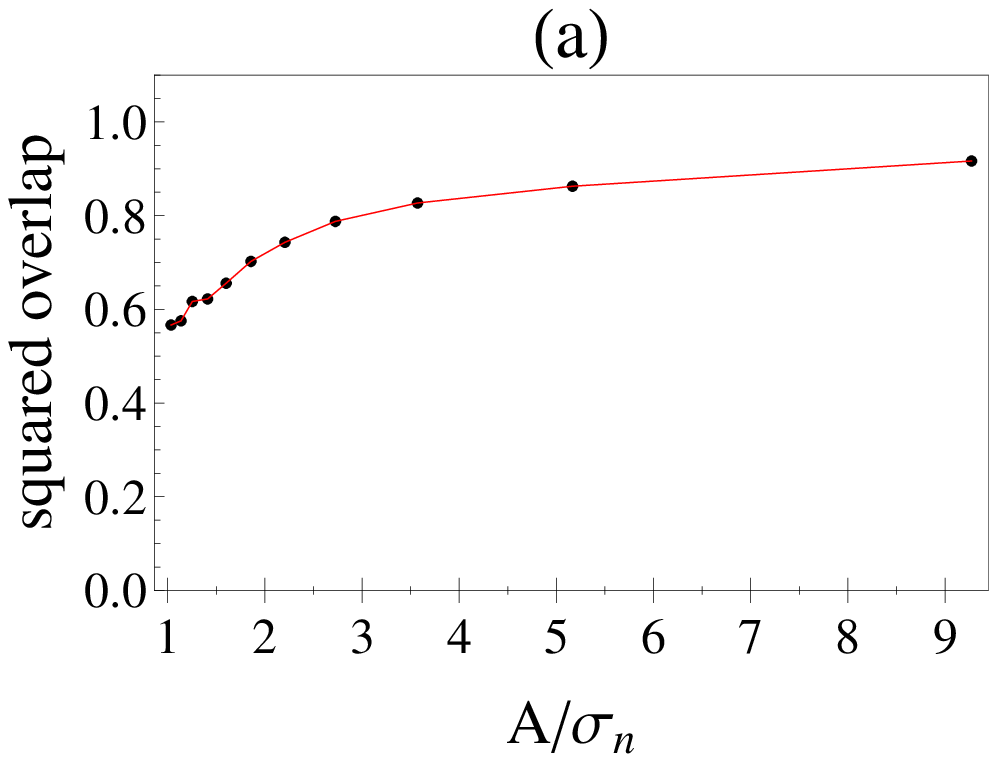}} 
\resizebox{0.45\columnwidth}{!}{\includegraphics{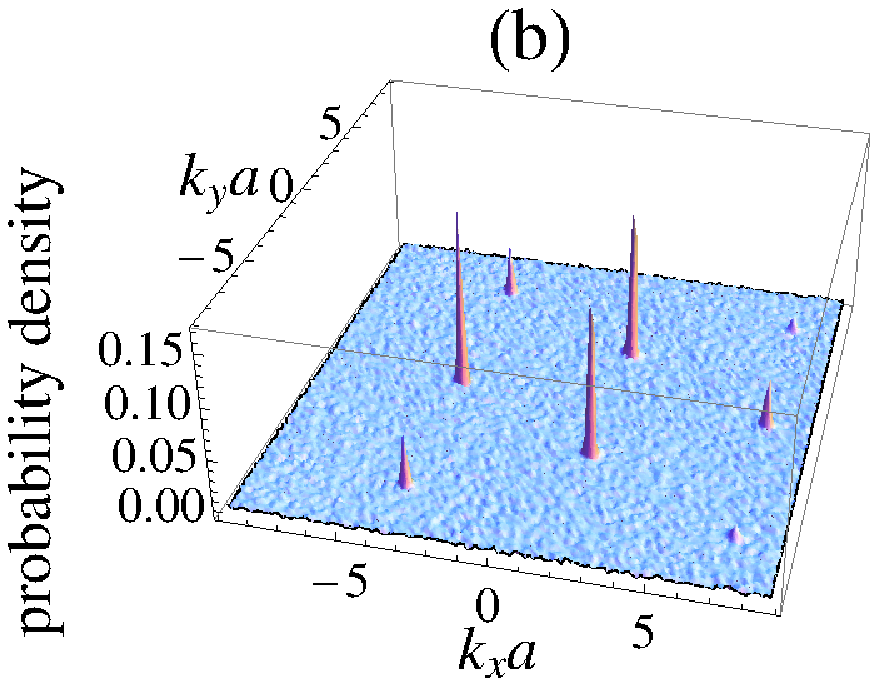}} 
\resizebox{0.5\columnwidth}{!}{\includegraphics{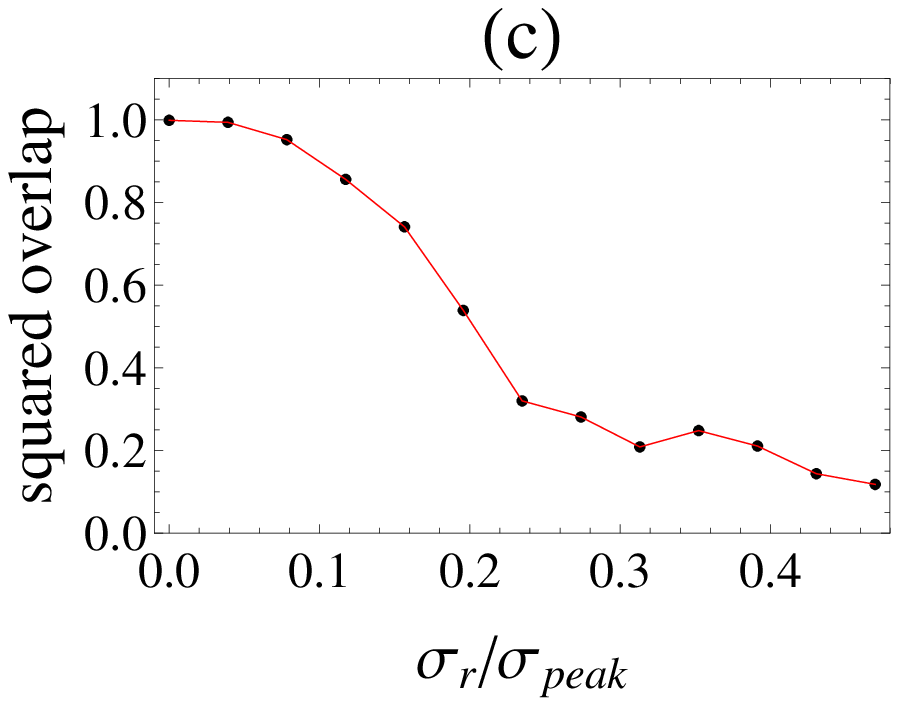}} 
\resizebox{0.45\columnwidth}{!}{\includegraphics{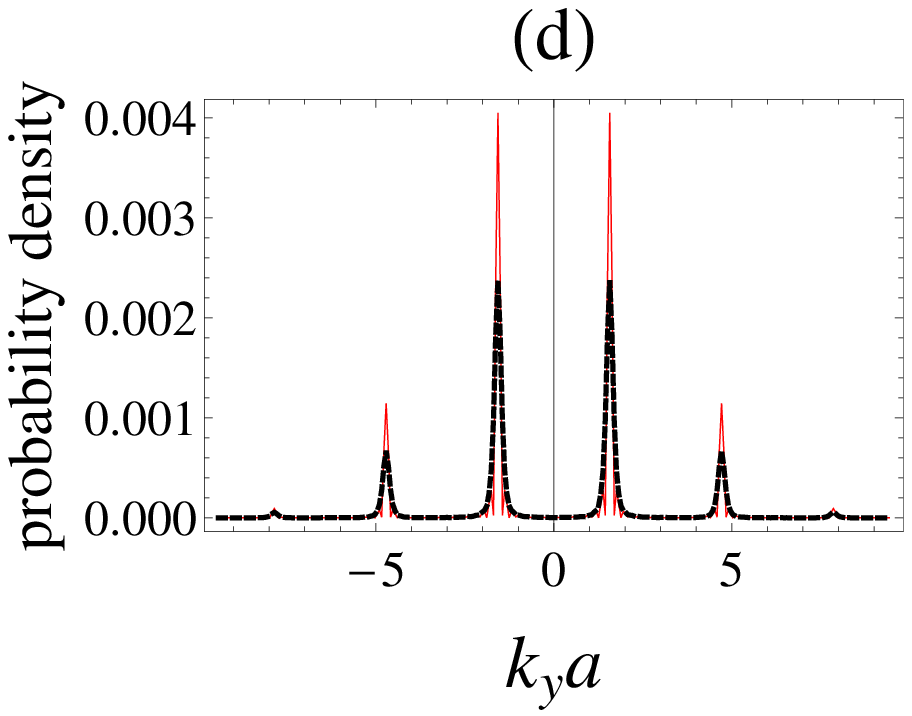}} 
\caption{
 Panel (a): the squared overlap between the model state $\psi_{\vect{k}_0}(\vect{r}_i)$ and the retrieved wave-function versus the signal to noise ratio $A/\sigma_n$, where $A$ is the average value of  $|\tilde\psi_{0}(\vect{\vect{k}})|^2$, calculated on a region where $k\le 2\pi/a$, and $\sigma_{n}$ is the standard deviation of Gaussian white noise which is added to the probability density. Panel~(b) shows an example of the noisy probability density for  $A/\sigma_n=3.57$. The squared overlap corresponding to the wave-function retrieved from this image is 0.83. Panel~(c): the dependence of the squared overlap between the model state $\psi_{\vect{k}_0}(\vect{r}_i)$ and the retrieved wave-function on the resolution of an imaging system. $\sigma_r$ - the standard deviation of a Gaussian distribution convoluted with the image of the model state. $\sigma_{peak}$ - the standard deviation of the Gaussian fit to the highest Bragg peak. In panel (d) there is an example of the convoluted image (black dashed line) for 
$\sigma_
r/\sigma_{peak}=0.12$ compared to the original image (red 
solid line).  For this example the squared 
overlap of the retrieved wave-function with the exact state is 0.83. The model wave-function used in the present analysis corresponds to the TF radius $R_{TF}=15 \,a$.
}
\label{overlaps}
\end{center}
\end{figure}

In the presence of a trap, even if the atomic density can not be measured in situ, it can be quite accurately estimated. In the case of an optical lattice potential combined with a shallow trap, the modulus of the condensate wave-function corresponding to the ground state of repulsively interacting atoms reads
\be
|\psi(\vect{r})| \approx |\varphi_{TF}(\vect{r})|\sum_i |w_0(\vect{r}-\vect{r}_i)|,
\label{trapden}
\ee
where $\varphi_{TF}(\vect{r})$ is the Thomas-Fermi (TF) envelope and $w_0(\vect{r}-\vect{r}_i)$ is the lowest band Wannier function localized at $\vect{r}_i$ \cite{exptech}. At low temperatures, the density fluctuations are suppressed due to the repulsive particle interactions \cite{lewenstein01,arlt03} and the density profile is identical to the ground state case, Eq.~(\ref{trapden}). This additional information can be used in the phase retrieval algorithms. Indeed, the projection on the support $S$ can be substituted with the projection on the modulus of the wave-function. That is, in the item (iii) described in the previous paragraph, at each iteration of the algorithm in order to get $\psi^{(n)}(\vect r)$ the modulus of the function obtained in the inverse Fourier transform is substituted with Eq.~(\ref{trapden}) \cite{supplement}.

So far we have assumed that images of atomic densities obtained after TOF correspond to the far field regime. This regime is rarely reached in an experiment where typically $t_{TOF}$ is about 20-30~ms. In a case of the finite expansion time of the atomic cloud released from an optical lattice potential, the density (\ref{intensity}) has to be modified \cite{Svistunov2008}, i.e. $I(\vect{r}) \propto |\tilde{w}_0(\vect{k})|^2 \left|\sum_{i} e^{i \vect{k}\cdot\vect{r}_i}\psi(\vect{r}_i) e^{-i\beta \vect{r}^2_i} \right|^2$ where $\tilde{w}_0(\vect{k})$ is the Fourier transform of the Wannier function and $\beta=m/(2\hbar t_{TOF})$. A new phase factor $e^{-i\beta \vect{r}^2_i}$ accounts for the deformation and widening of Bragg peaks and can be easily included in our phase retrieval algorithm.  Indeed, at each iteration of the algorithm after the inverse Fourier transform is applied, one has to perform an additional transformation $\psi^{(n)}(\vect{r})\rightarrow e^{i\beta \vect{r}^2}\psi^{(n)}(\vect{r})$ only \
cite{
supplement}. The initial stage of atomic gas expansion is influenced also by the atomic interactions. This effect is weaker than the near field corrections \cite{Svistunov2008}, nevertheless, it can be taken into account by a 
suitable choice the 
Wannier functions so that the envelope $|\tilde{w}_0(\vect{k})|^2$ of the image is properly reproduced.

To summarize, an experimentally measured image after the TOF, theoretically estimated density $|\psi(\vect{r})|^2$ and its support $S$ are information used in our phase retrieval algorithm which allows us to find unknown phases $\varphi(\vect{r})$ and consequently obtain the entire wave-function $\psi(\vect{r})=|\psi(\vect{r})|e^{i \varphi(\vect{r})}$. In ref. \cite{supplement} we describe in detail how the information about $|\psi(\vect{r})|^2$ and its support $S$ is used in order to obtain the results presented in the following.

Let us apply the described phase microscopy to a two-dimensional (2D) problem of a BEC in a triangular optical lattice which can be described by the Bose-Hubbard Hamiltonian \cite{struck2011,sacha2012,struck2012,struck2012a,kosior2013,struck2013}. Primitive vectors of the Bravais lattice read $\vect{a}_1=a\vect{e}_x$ and $\vect{a}_2=a(\sqrt{3}\vect{e}_y+\vect{e}_x)/2$ where $a$ is the lattice constant. The presence of a shallow harmonic trap is also assumed. The periodic shaking of the lattice allows for the modification of tunneling amplitudes of the Bose-Hubbard model \cite{eckardt05,eckardt2010,struck2011,struck2012,arimondo12,struck2013}. The modification of the kinetic part of the Hamiltonian indicates the presence of a gauge vector potential \cite{struck2012,struck2012a}. Especially for the negative tunneling amplitudes we deal with staggered magnetic fluxes. The corresponding dispersion relation reveals two degenerate non-equivalent minima at $\pm\vect{k}_0$ where $\vect{k}_0a=4\pi\vect{e}_x/3$.
 The degenerated ground states of the system can be written as $\psi_{\pm\vect{k}_0}(\vect{r})=\varphi_{TF}(\vect{r})\left(\sum_i e^{\pm i \vect{k}_0 \vect{r}_i} w_0(\vect{r}-\vect{r}_i)\right)$ where we approximate the Wannier functions by 2D Gaussian distributions with the standard deviation $\sigma_{W}=0.155\,a$. The non-zero quasi-momentum vectors of the ground states imply that the order parameter is complex.  

 Assume that atomic gas, prepared in the ground state $\psi_{\vect{k}_0}(\vect{r})$, performs the TOF expansion and its density is measured in the far field regime, i.e. $\beta R_{TF}^2=mR_{TF}^2/(2\hbar t_{TOF})\approx 0$. Figure~\ref{overlaps} illustrates the influence of experimental imperfections on the accuracy of the phase microscopy. After adding Gaussian white noise to the {\it detected} image, the overlap of the retrieved wave-functions with the exact ground state is diminished (the overlaps are calculated between wave-functions which are firstly projected on the Wannier basis vectors). Fig.~\ref{overlaps}a shows how the overlap changes with signal to noise ratio $A/\sigma_n$, where $A$ is the average value of  $|\tilde\psi_{0}(\vect{\vect{k}})|^2$, calculated on a region where $k\le 2\pi/a$, and $\sigma_{n}$ is the standard deviation of the noise distribution. Each point of the plot has been obtained by running the phase retrieval algorithm for 30 sets of 
initially random phases $\varphi^{(0)}(\vect{r})$ 
and then the best result is chosen  (i.e. the result $\tilde\psi(\vect k)$ with minimal value of $Q=\int d^2k||\tilde\psi(\vect k)|^2-|\tilde\psi_0(\vect k)|^2|$). We see that even substantial amount of noise allows one to retrieve 
the wave-function with the reasonable overlap. In Fig.~\ref{overlaps}b we present an example of a noisy image corresponding to $A/\sigma_{n}=3.57$ which leads to the squared overlap of 0.83. 

The finite resolution of an imaging system is another important source of distortion of experimental density images. To analyze its impact on the phase microscopy we have convoluted the previously considered model image with the Gaussian distribution. Standard deviation $\sigma_r$ of the distribution is a measure of the resolution. In Fig.~\ref{overlaps}c we show the squared overlap of the retrieved wave-functions with the exact state versus $\sigma_r$. The overlap drops quite quickly which seems to be not very surprising because broadening of the Bragg peaks reduces artificially the coherence length of the system. 

\begin{figure}[ht]
\begin{center}
\resizebox{0.49\columnwidth}{!}{\includegraphics{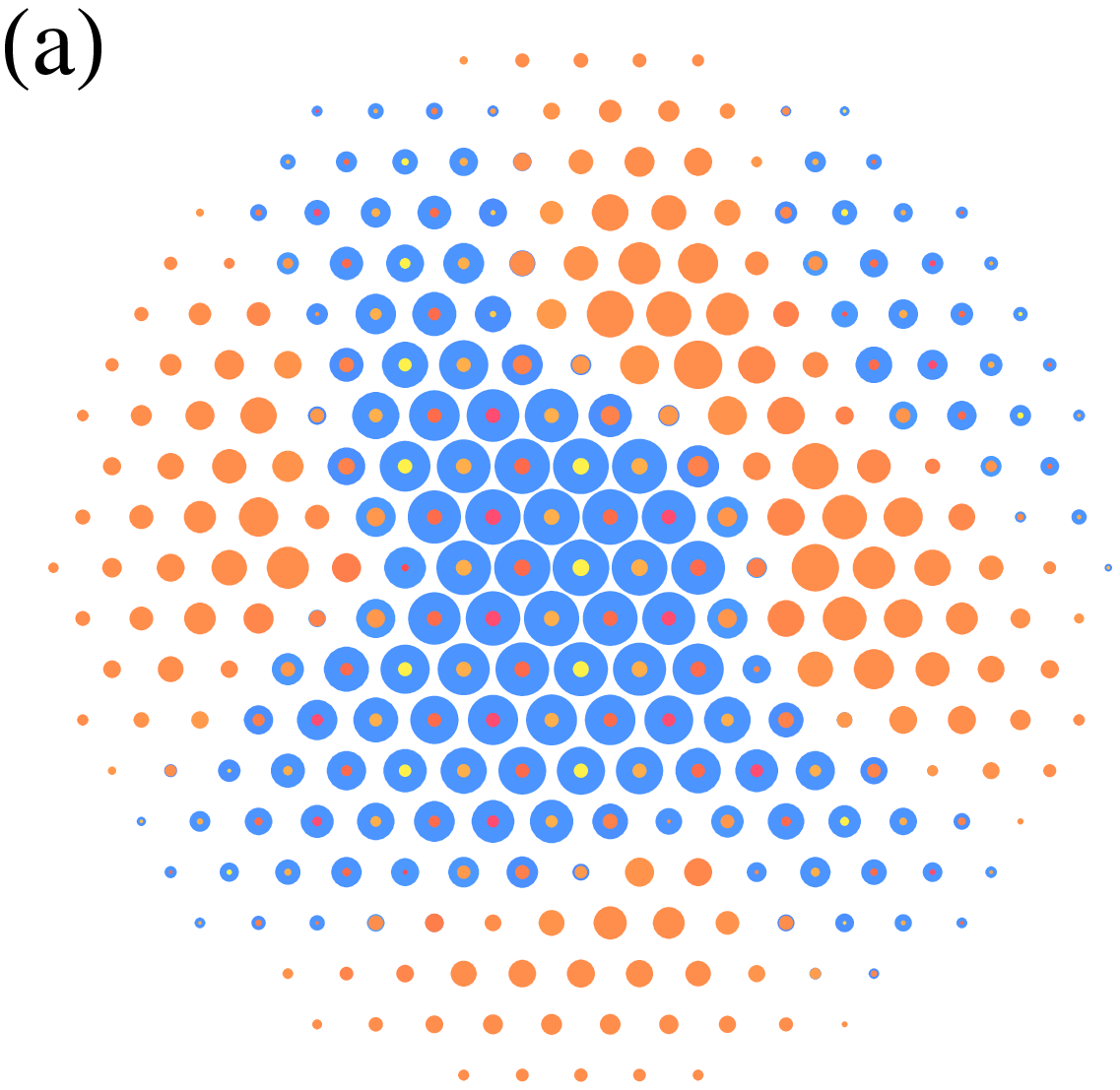}} 
\resizebox{0.49\columnwidth}{!}{\includegraphics{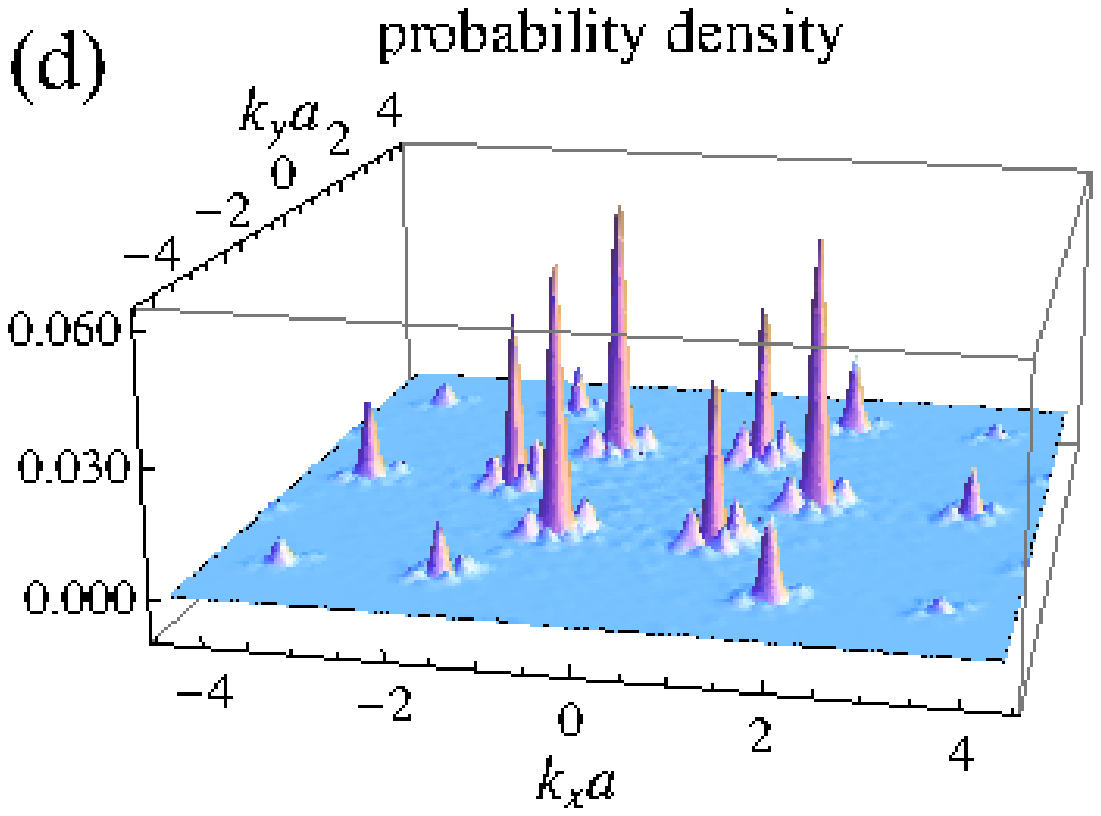}} 
\resizebox{0.49\columnwidth}{!}{\includegraphics{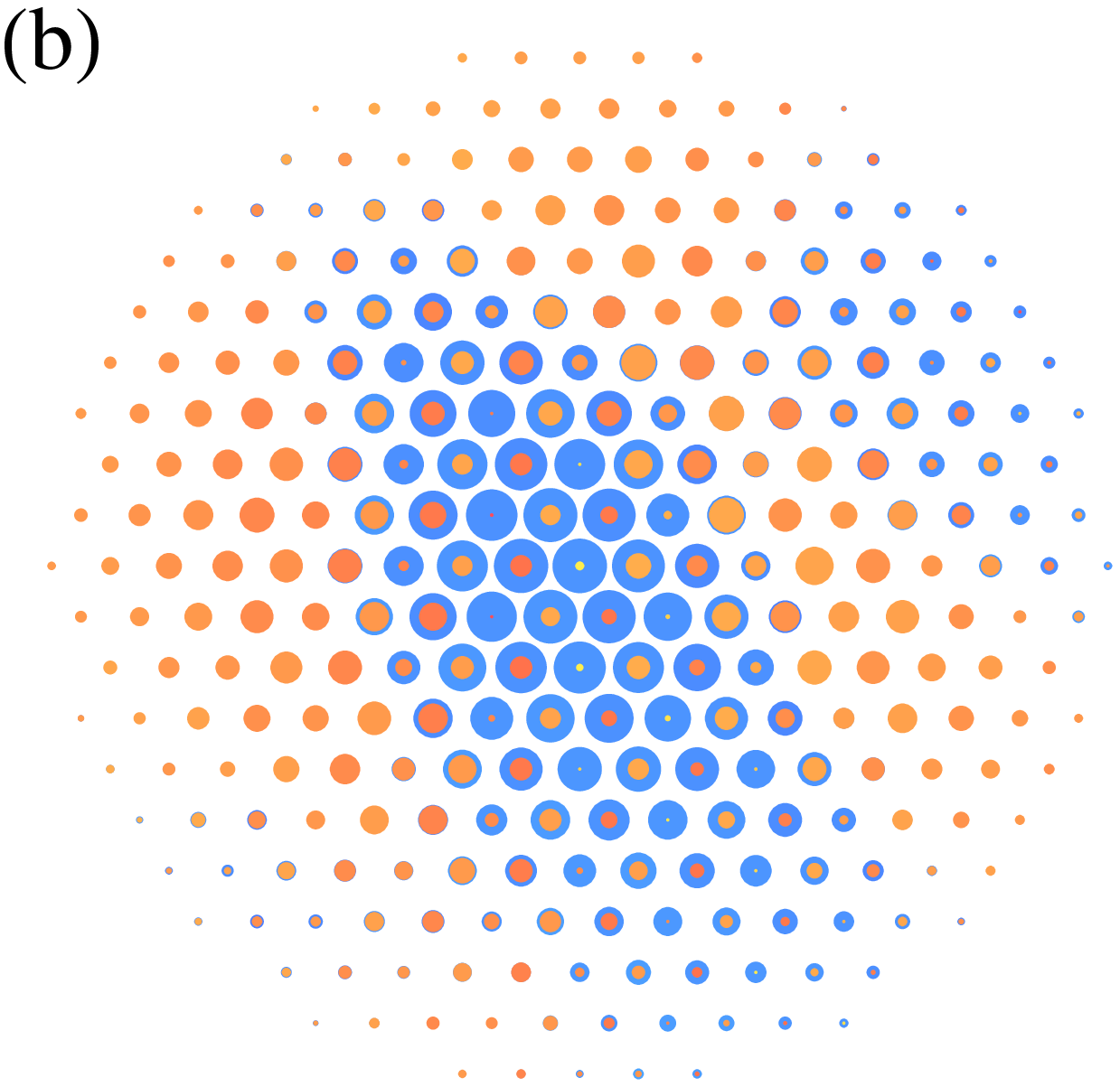}} 
\resizebox{0.49\columnwidth}{!}{\includegraphics{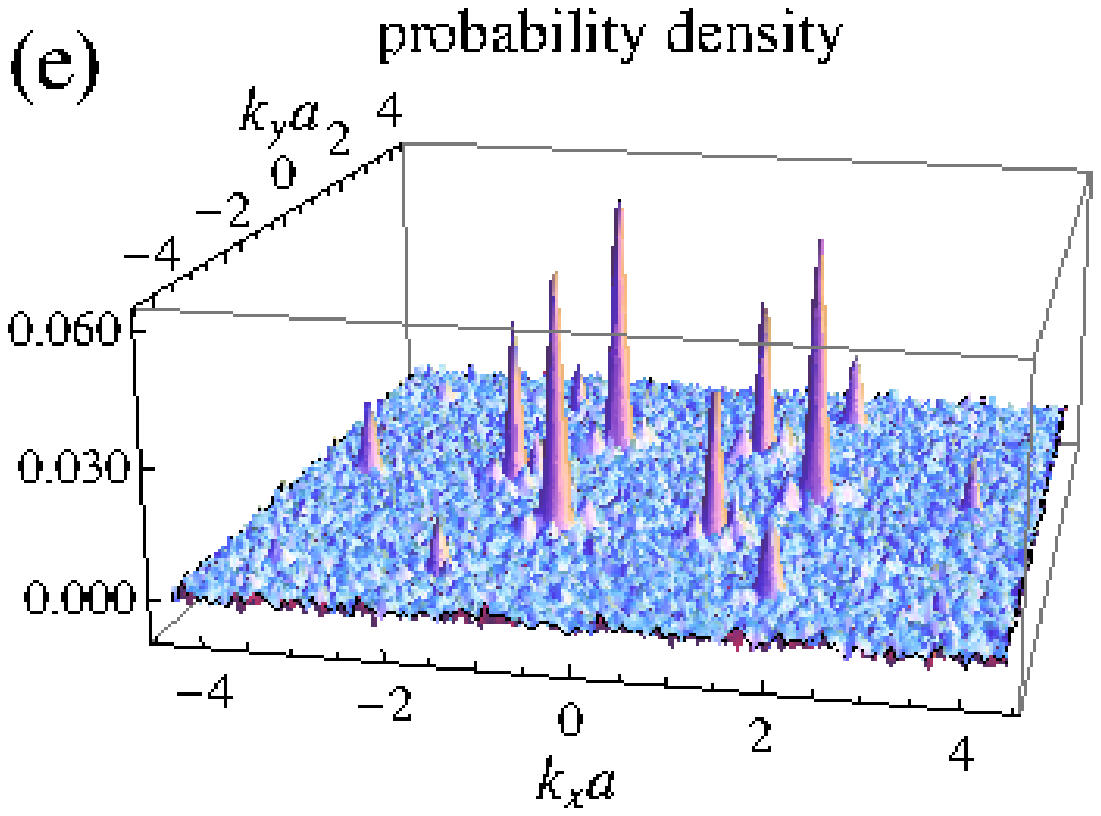}} 
\resizebox{0.49\columnwidth}{!}{\includegraphics{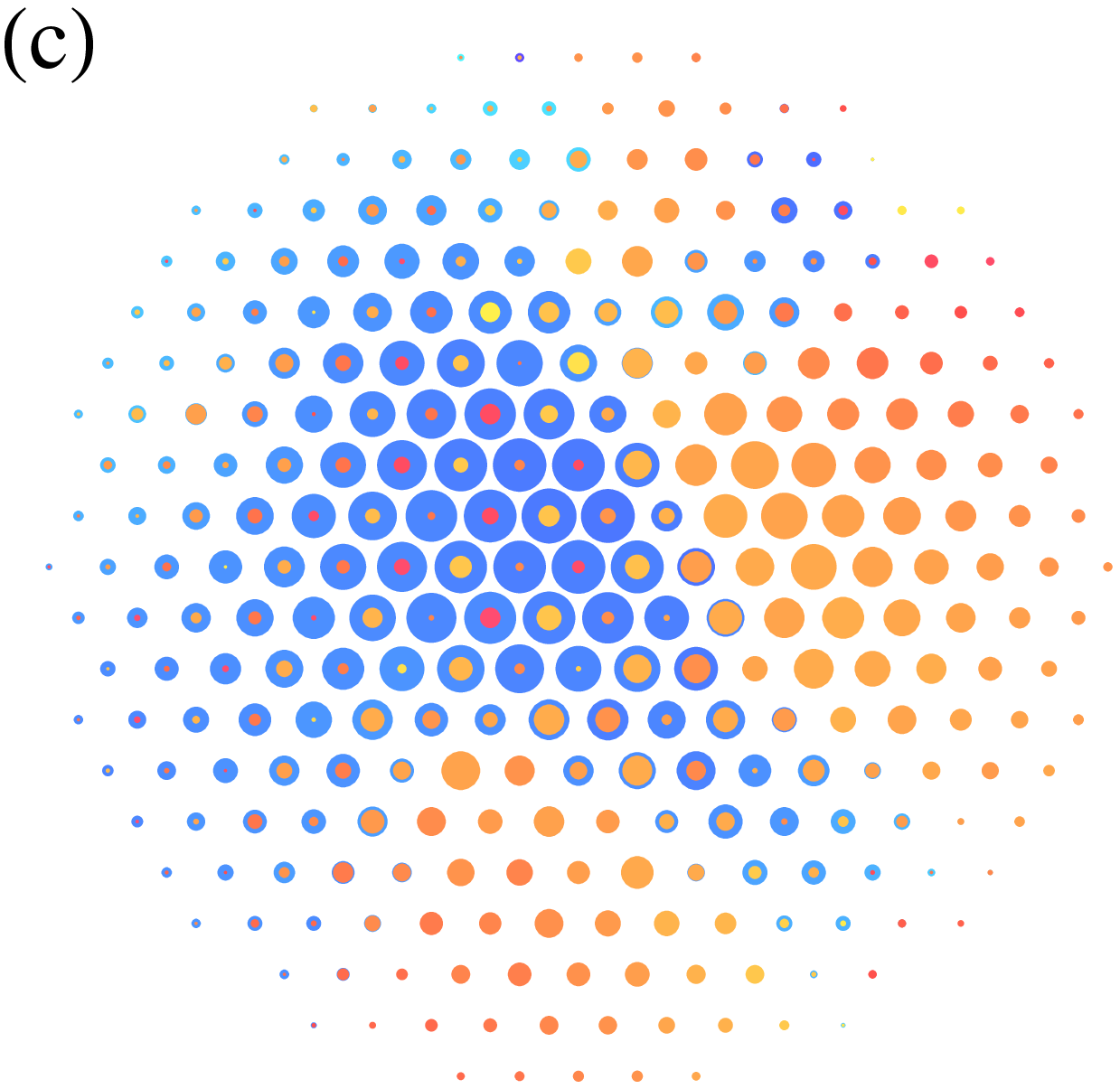}} 
\resizebox{0.49\columnwidth}{!}{\includegraphics{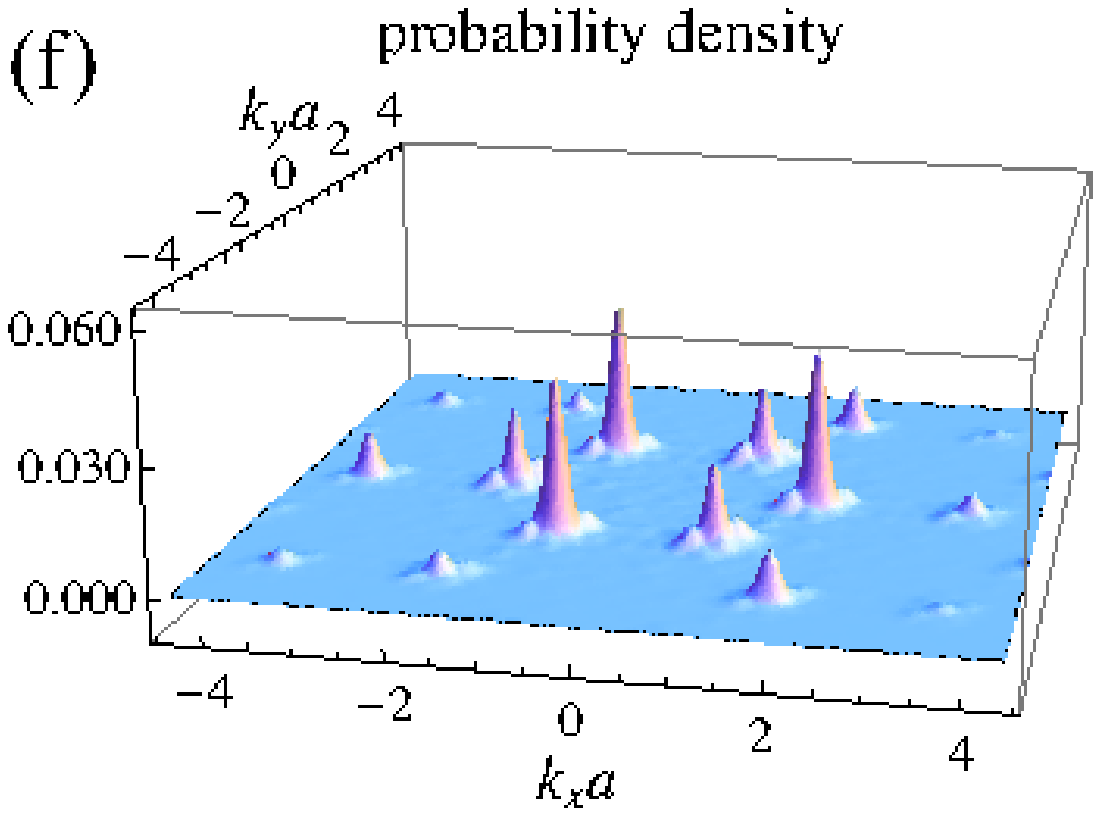}} 
\resizebox{0.45\columnwidth}{!}{\includegraphics{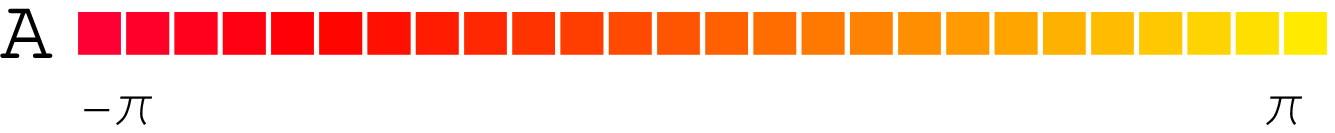}} 
\resizebox{0.45\columnwidth}{!}{\includegraphics{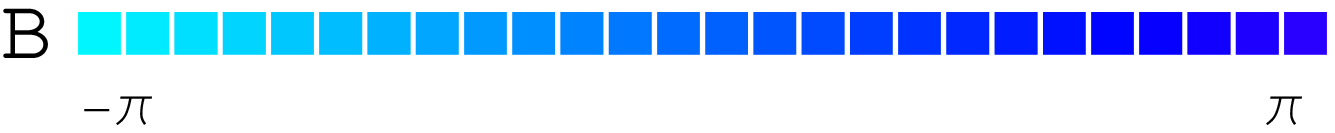}} 
\caption{ (Color online) Left column: plots of $\overline{\alpha}_{\pm\vect{k}_0}(\vect{r}_i)$, Eq.~(\ref{aalpha}), obtained on the basis of the exact model state  (a) and on the basis of the retrieved wave-functions (b) and (c). Size of a dot at a given lattice site $\vect{r}_i$ is proportional to $|\overline{\alpha}_{\pm\vect{k}_0}(\vect{r}_i)|$. Warm colors indicate complex phase of $\overline{\alpha}_{\vect{k}_0}$, i.e. regions A, while cold colors show complex phase of $\overline{\alpha}_{-\vect{k}_0}$, i.e. regions B. Right column: the momentum distribution corresponding to the model wave-function (d), the momentum distribution with additional white noise where the signal to noise ratio $A/\sigma_n=0.5$ (e) and the distribution convoluted with a Gaussian distribution with the standard deviation $\sigma_r=0.1\sigma_{peak}$ (f).
The TF radius of the model wave-function $R_{TF}=10 \,a$. The squared overlap of the retrieved wave-functions with the exact state are 0.6 (b) and 0.3 (c).}
\label{modeldomain} 
\end{center}
\end{figure}

The phase microscopy can be applied even if experimental images do not correspond to the far-field regime because the important additional phase factor $e^{-i\beta \vect{r}_i^2}$ can be easily included in the algorithm \cite{supplement}. However, the presence of this phase factor makes the basin of attraction of a desired solution smaller. Therefore, the algorithm has to be run for greater number of randomly chosen phases $\varphi^{(0)}(\vect{r})$ in order to find the desired wave-function.

 The performed analysis of experimental imperfections allows us to estimate requirements that have to be fulfilled in order to apply the phase microscopy. Assume that experimental images are obtained in the far-field limit. According to Fig.~\ref{overlaps}, the resolution of an imaging system has to fulfill $\sigma_r<0.1\sigma_{peak}$, where $\sigma_{peak}\approx 2 /R_{TF}$, so that the retrieved wave-function has the squared overlap with the exact solution greater than 0.8. In the near-field regime the Bragg peaks can be much wider than $2/R_{TF}$, but then they may possess small additional structures superimposed on the bell profiles which must not be blurred by an imaging system. Therefore, the requirement for the resolution should not be weakened in the near-field regime. The presence of noise additionally worsens accuracy of the retrieved wave-function.  However, the discrimination of the noisy image with the threshold of $3\sigma_n$ (i.e. $|\tilde\psi_0(\vect k)|^2$ is set 0 if it is smaller than $3\
sigma_n$) can 
help significantly. For example, we have checked that for $\sigma_r/\sigma_{
peak}=0.2$, if we apply the discrimination in the noisy image ($A/\sigma_n=3$), the  overlap is smaller by 0.1 only as compared to the case without the noise.

The phase microscopy is very useful in detection and visualization of spatial domains. In order to demonstrate it we have prepared a wave-function which consists of 5 spatially separated regions where either $\psi_{\vect{k}_0}$ or $\psi_{-\vect{k}_0}$ is present. To visualize domains corresponding to the $\psi_{\vect{k}_0}$ and $\psi_{-\vect{k}_0}$ states let us define quantities $\alpha_{\pm\vect{k}_0}(\vect{r}_i)= \psi^*_{\pm\vect{k}_0}(\vect{r}_i) \psi(\vect{r}_i)$.  In the egions dominated by $\psi_{\vect{k}_0}$ (let us call them regions A) we should get, e.g., $\alpha_{\vect{k}_0}(\vect{r}_i)\propto|\varphi_{TF}(\vect{r}_i)|^2e^{i\zeta}$ where $\zeta$ is a constant global 
phase. However, in the regions where the state $\psi_{-\vect{k}_0}$ exists (regions B), $\alpha_{\vect{k}_0}(\vect{r}_i)\propto|\varphi_{TF}(\vect{r}_i)|^2e^{i\xi}e^{-i2\vect{k}_0\cdot\vect{r}_i}$ where $\xi$ is a 
constant phase. Let us now average $\alpha_{\pm\vect{k}_0}$ over a small volume,
\begin{eqnarray}
\overline{\alpha}_{\pm\vect{k}_0}(\vect{r}_i)=\frac{1}{7}{\sum_j}'\alpha_{\pm\vect{k}_0}(\vect{r}_j),
\label{aalpha}
\end{eqnarray}
where ${\sum_j}'$ denotes the sum over $j=i$ and the nearest neighbors of the $i$-site (one may consider also averaging extended to next to the nearest neighbors or even further). In regions A we expect $\overline\alpha_{\vect{k}_0}(\vect{r}_i)\approx \alpha_{\vect{k}_0}(\vect{r}_i)$ and $\overline\alpha_{-\vect{k}_0}(\vect{r}_i)\approx 0$ while in regions B $\overline\alpha_{\vect{k}_0}(\vect{r}_i)\approx 0$ and $\overline\alpha_{-\vect{k}_0}(\vect{r}_i)\approx \alpha_{-\vect{k}_0}(\vect{r}_i)$.  In Fig.~\ref{modeldomain}a we show $\overline\alpha_{\pm\vect{k}_0}(\vect{r}_i)$ calculated for the exact model state whose momentum distribution is presented in Fig.~\ref{modeldomain}d. Five spatially separated domains of different sizes are clearly visible in Fig.~\ref{modeldomain}a. If we add substantial noise ($A/\sigma_n=0.5$) to the momentum distribution, see Fig.~\ref{modeldomain}e, the phase retrieval algorithm results in the domain picture depicted in Fig.~\ref{modeldomain}b. The noisy image has been 
discriminated with the threshold $3 \sigma_n$, before the algorithm is used. When the original momentum distribution is convoluted with a Gaussian distribution characterized by the standard deviation $\sigma_r=0.1\sigma_{peak}$, see Fig.~\ref{modeldomain}f, the phase microscopy leads to the domain picture shown in Fig.~\ref{modeldomain}c. Comparing Figs.~\ref{modeldomain}a-c one can see that despite the considerable imperfections assumed in the initial data, the largest domains are quite well reproduced by the phase retrieval algorithm. Small domains may either glue together (Fig.~\ref{modeldomain}b) or disappear (Fig.~\ref{modeldomain}c) especially if they are localized on the border where the probability density is very small. In order to estimate the areas occupied by the different kinds of the domains we have calculated the squared overlap of the exact model wave-function with the ground states $\psi_{\vect k_0}$ and $\psi_{-\vect k_0}$ 
which turn out to be 0.35 and 0.16, respectively. In a case 
of the noisy image the corresponding overlaps are 0.5 and 0.26 while for the convoluted image we obtain 0.28 and 0.11, respectively. These numbers illustrate the accuracy of the phase microscopy in determination of domain sizes in the case when substantial experimental imperfections are present.

To conclude, we have demonstrated that the phase of a Bose-Einstein condensate wave-function of ultra-cold atomic gases in optical lattice potentials in 2D can be detected. Knowledge of an atomic density after time-of-flight expansion and a theoretical estimate for the initial atomic density in the presence of an external potential are used in phase retrieval algorithms. We analyze influence of experimental imperfections on such a phase microscopy. It seems that the most important requirement is the sufficiently high resolution of an imaging system. The phase microscopy is particularly useful in situations when the order parameter of a system is complex,  for example in the presence of artificial gauge potentials \cite{struck2011,spielman12} or in multi-orbital superfluid phase in optical lattices \cite{hemmerich11, sengstock12,olschlager2013}. Additionally, the phase microscopy can be applied to observe long range phase fluctuations. In the present letter we concentrate on a BEC in a triangular optical 
lattice subjected to artificial gauge fields. The phase microscopy allows for visualization of spatial domains whose presence is discussed in the literature \cite{struck2011,struck2013}.  Recently in Ref.~\cite{parker13} a different method for 
visualization of Weiss domains has been proposed. In this method, however, phase relation between domains is lost. On the contrary, the method presented in this letter allows one to retrieve all phase informations. 

 We are grateful to: J. Struck, M. Weinberg, C. \"Olschl\"ager, P. Windpassinger, J. Simonet, and K. Sengstock  for fruitful discussions and useful comments during preparation of the manuscript.
We acknowledge support of Polish National Science Center via project DEC-2012/04/A/ST2/00088. AK acknowledges support in a form of a special scholarship of Marian Smoluchowski Scientiﬁc Consortium
“Matter Energy Future” from KNOW funding.


\section*{Supplemental material}
Knowledge of the Fourier transform of a wave-function $\psi(\vect{r}_i)$, i.e.
\be
\tilde\psi_0(\vect{k})\propto \sum_i e^{i \vect{k}\cdot\vect{r}_i}\psi(\vect{r}_i), 
\label{intensityS}
\ee
 is sufficient in order to obtain $\psi(\vect{r}_i)$. However, if only the absolute value $|\tilde\psi_0(\vect{k})|$ is known, the information on the phase is lost and the inverse Fourier transform cannot be applied. Phase retrieval algorithms allows one to recover the phase provided additional constrains on a wave-function can be imposed. Assume that the support $S$ of $\psi(\vect{r}_i)$ is known and it is finite, that is usually the case for trapped atomic gases  ($S$ should be chosen so that the probability of the detection of an atom outside $S$ is negligibly small).  Then, apart from the fact that the sought wave-function belongs to a set of all functions with the same modulus of its Fourier transform $|\tilde{\psi}_0(\vect{k})|$, it also belongs to a set of functions with the same support $S$. The phase retrieval algorithms seek for the intersection of these two sets, where $\psi(\vect{r}_i)$ is located \cite{Fienup1987_S,miaosayre2000_S,marchesini2007_S}. 

In the error reduction (ER) algorithm successive projection of an estimate of a wave-function onto the sets allows one to find the desired solution \cite{marchesini2007_S}. That is, starting with, e.g., randomly chosen phases ${\varphi}^{(0)}(\vect{r})$ of $\psi(\vect{r})$, the following iteration is performed
\be
\psi^{(n+1)}=P_SP_M \psi^{(n)},
\ee
where
\be
P_S \psi^{(n)}(\vect{r})= 
\left\{
\begin{array}{ll}
 \psi^{(n)}(\vect{r}), & \mbox{if }\vect{r}\in S \\
 0, & \mbox{otherwise}
\end{array}
\right.,
\ee
and
\begin{eqnarray}
\label{invFFT}
P_M \psi^{(n)}(\vect{r}) &=&  \mathcal{F}^{-1} \left( |\tilde{\psi}_0(\vect{k})| e^{i \tilde{\varphi}^{(n)}(\vect{k})}\right),\\
\tilde{\varphi}^{(n)}(\vect{k})&=&\arg \tilde{\psi}^{(n)}(\vect{k}).\nonumber
\end{eqnarray}
The symbol $\mathcal{F}^{-1}$ denotes the inverse Fourier transform operator.  Ideally, the iteration of the ER algorithm should converge to $\tilde\psi(\vect k)$ for which the quantity 
\be
Q=\int d^2k||\tilde\psi(\vect k)|^2-|\tilde\psi_0(\vect k)|^2|,
\label{Q}
\ee 
equals zero. However, it is known that the ER algorithm tends to stuck at a local minimum of $Q$ and the hybrid input-output (HIO) algorithm, where 
\be
 \psi^{(n+1)}(\vect{r})=
\left\{
\begin{array}{ll}
 P_M\psi^{(n)}(\vect{r}), & \mbox{if }\vect{r}\in S \\
 \left(id -\eta P_M)\psi^{(n)}(\vect{r}\right), & \mbox{otherwise}
\end{array}
\right.,
\ee
is more efficient \cite{marchesini2007_S}. The parameter $\eta$ is usually chosen between $0.7$ and $0.9$. In practice the best convergence is achieved when after every 20 iterations of the HIO algorithm there is an iteration of the ER one, see \cite{marchesini2007_S}. 

 It is possible to construct simple homometric examples of $|\tilde\psi_0(\vect k)|^2$ that correspond to different non-equivalent $\psi(\vect r)$ (i.e. wave-functions which are not related to each other by e.g. a change of a global phase or origin shift). However, for more complex structures the non-uniqueness is very unlikely in 2D and higher dimensions and does not occur in practise, see discussion in \cite{marchesini2007_S}.

For very low temperatures, density fluctuations are strongly suppressed in trapped atoms with repulsive interactions \cite{lewenstein01_S,arlt03_S}. Then, the modulus of a wave-function $|\psi(\vect{r})|$ can be quite accurately estimated if parameters of an external trapping potential and number of atoms are known. This information can be used in the phase retrieval algorithms. Indeed, the support projection operator $P_S$ in the ER algorithm can be substituted with the projection 
\be
P_{|\psi|} \psi^{(n)}= |\psi(\vect{r})| e^{i \varphi^{(n)}(\vect{r})},
\ee
where 
\be
\varphi^{(n)}(\vect{r})=\arg \psi^{(n)}(\vect{r}).
\ee
Such a modified ER procedure is applied every 20 iterations of the HIO algorithm. 

If in an experiment atomic densities are measured after very long time-of-flight, the relation between an initial wave-function in the presence of an optical lattice and a measured image reduces to the Fourier transform Eq.~(\ref{intensityS}), where $\vect{k}=m \vect{r}/(\hbar t_{TOF})$. However, if $t_{TOF}$ is not very long, the near-field corrections have to be taken into account, i.e. 
\be
|\tilde{\psi}_0(\vect{k})|^2 \propto \left|\sum_{i} e^{i \vect{k}\cdot\vect{r}_i}\psi(\vect{r}_i) e^{-i\beta \vect{r}^2_i} \right|^2,
\ee
where $\beta=m/(2\hbar t_{TOF})$ \cite{Svistunov2008_S}. The presence of the phase factor $e^{-i\beta \vect{r}^2_i}$ can be easily included in the phase retrieval algorithm. Indeed, after the inverse Fourier transform Eq.~(\ref{invFFT}) is applied, one has to perform an additional transformation $\psi^{(n)}(\vect{r})\rightarrow e^{i\beta \vect{r}^2}\psi^{(n)}(\vect{r})$ only.

The phase retrieval algorithm may stuck at local minimum. Therefore, in order to find the desired solution, the algorithm should be initiated many times with different, randomly chosen, phases ${\varphi}^{(0)}(\vect{r})$. In the present publication, the algorithm has been initiated 30-100 times. We have observed that in the near-field regime more extensive sampling has to be used than in the far-field limit.


\end{document}